\documentclass[11pt]{article}
\usepackage{amssymb}
\usepackage{graphics}
\usepackage{epsfig}
\usepackage{a4wide}

\newcommand{\eezzh}{$e^+e^- \to Z^0Z^0H^0~$}
\newcommand{\eehhz}{$e^+e^- \to H^0H^0Z^0~$}
\newcommand{\eezh}{$e^+e^- \to Z^0H^0~$}
\newcommand{\nb}{\nonumber}

\begin{document}

\title{ Electroweak radiative corrections to the Higgs-boson production in association
with $Z^0$-boson pair at $e^+ e^-$ colliders \footnote{Supported
by National Natural Science Foundation of China.}} \vspace{3mm}

\author{{\small{Zhou Ya-Jin$^{2}$, Ma Wen-Gan$^{1,2}$, Zhang Ren-You$^{2}$, Jiang Yi$^{2}$, and Han Liang$^{2}$}}\\
{\small $^{1}$CCAST (World Laboratory), P.O.Box 8730, Beijing, 100080, People's Republic of China} \\
{\small $^{2}$ Department of Modern Physics, University of Science and Technology}\\
{\small of China (USTC), Hefei, Anhui 230026, P.R.China}}
\date{}
\maketitle \vskip 8mm

\begin{abstract}
We present the full ${\cal O}(\alpha_{{ew}})$ electroweak
radiative corrections to the Higgs-boson production in association
with $Z^0$-boson pair at an electron-positron linear collider(LC)
in the standard model. We analyze the dependence of the full
one-loop corrections on the Higgs-boson mass $m_{H}$ and colliding
energy $\sqrt{s}$. We find that the corrections significantly
suppress the Born cross section, and the ${\cal O}(\alpha_{{ew}})$
electroweak radiative corrections are generally between $1.0\%$
and $-15\%$ in our chosen parameter space, which should be taken
into consideration in the future precise experiments.
\end{abstract}

\vskip 6cm {\large\bf PACS: 12.15.Lk, 14.80.Bn,
14.70.Hp,11.80.Fv}\\
 \vfill \eject
\baselineskip=0.34in

\par
\section{Introduction}

\par
One of the most important missions of the future high energy
experiments is to search for scalar Higgs-boson, which is believed
to be responsible for the breaking of the electroweak symmetry and
the generation of masses for the fundamental particles in the
standard model(SM)\cite{sm,higgs}. Until now the Higgs-boson
hasn't been observed yet, only LEP II experiment provides a lower
bound of 114.4 GeV\cite{LEPh1} and an upper bound of 260
GeV\cite{LEPh2} for the mass of the SM Higgs-boson at $95\%$
confidence level. People believe that with the help of future high
energy colliders, such as the CERN large hadron collider(LHC), and
the linear colliders, TESLA, NLC, JLC and CERN CLIC, the existence
of the Higgs-boson would be proved or excluded in the experiments.

\par
As far as we know, the present precise experimental data have
shown an excellent agreement with the predictions of the SM except
the Higgs sector\cite{SM experiments}. These data strongly
constrain the couplings of the gauge-boson to fermions
($g_{Zf\bar{f}}$ and $g_{Wf\bar{f^{\prime}}}$), and gauge bosons
self-couplings, but say little about the couplings between
Higgs-boson and gauge bosons, which wouldn't exist if the
corresponding scalar field has no vacuum expectation value. In
order to reconstruct Higgs potential, the precise predictions for
Higgs couplings, which include Yukawa couplings, the couplings of
Higgs to gauge bosons and the Higgs self-couplings, are necessary.
At an $e^+e^-$ linear collider with $\sqrt{s}\simeq 300-500~GeV$,
Higgs-boson with an intermediate mass value would be produced
mainly via the Higgs strahlung process \eezh, and the coupling of
Higgs-boson to $Z$-bosons is probed best in the measurement of the
cross sections of the Higgs strahlung process \eezh and the
$WW/ZZ$ fusion processes $e^+e^- \to H^0 \nu \bar{\nu}$ and
$e^+e^- \to H^0 e^+e^-$. In Refs.\cite{Precision}, it shows that
the coupling $g_{ZZH}$ can be determined at a few percent level
for a $120~GeV$ Higgs-boson with an integrated luminosity of
$500~fb^{-1}$ from the production cross section through the
process \eezh. There is another class of processes which is
interesting for the studies of Higgs physics at linear collider
called Higgs-boson production in association with a pair of final
particles which can be used to test the Yukawa couplings,
couplings between Higgs-boson and gauge bosons, and Higgs
self-couplings. For example, \eezzh process is not only an
important process in probing $g_{ZZH}$, but also possible to
provide further tests for the quadrilinear couplings(such as
C-violating $HZZZ$ or $H\gamma ZZ$), which do not exist at
tree-level in the SM, because these quadrilinear couplings would
induce deviations from the SM predicted observables\cite{eehvv}.
We believe that once the neutral Higgs-boson is discovered and its
mass is determined, the double $Z^0$-bosons production through
\eezzh process may provide the detail information of the coupling
between Higgs-boson and $Z^0$ gauge bosons, which directly
reflects the role of the Higgs-boson in electroweak symmetry
breaking. Moreover, a theoretical accurate estimate of this class
of processes is essential, since \eezzh process could be potential
backgrounds for possible new physics.

\par
There have been already some theoretical works in investigating
the Yukawa couplings, the Higgs self-couplings and Higgs couplings
with gauge boson pair in the SM at LC, for example, the
calculations of the NLO QCD and one-loop electroweak corrections
to the $e^+e^- \to t \bar{t} H^0$ process in
Refs.\cite{ee1,youyu,belanger,ljj} and $\gamma\gamma\to
t\bar{t}H^0$ process in Ref.\cite{chenhui} in probing Yukawa
coupling, the one-loop electroweak corrections to the process
\eehhz in Ref.\cite{ZhangRY,Belanger} for testing Higgs
self-coupling. The Higgs productions in association with vector
gauge bosons($e^+e^- \to H^0 W^+W^-$, $e^+e^- \to H^0Z^0Z^0$ and
$e^+e^- \to H^0Z^0 \gamma$) in the SM for testing the couplings
between Higgs-boson and gauge bosons were studied at the
tree-level in Ref.\cite{eehvv}. In this work we calculate the full
one-loop electroweak corrections to the process \eezzh in the SM.
The paper is arranged as follows: In Section II we give the
analytical calculations of the Born cross section and the full
${\cal O}(\alpha_{ew})$ electroweak corrections to the \eezzh
process. In Section III we present some numerical results, and
finally a short summary is given.

\par
\section{Calculation of \eezzh}

\par
\begin{figure}[htbp]
\vspace*{-0.3cm} \centering
\includegraphics*[130pt,412pt][560pt,510pt]{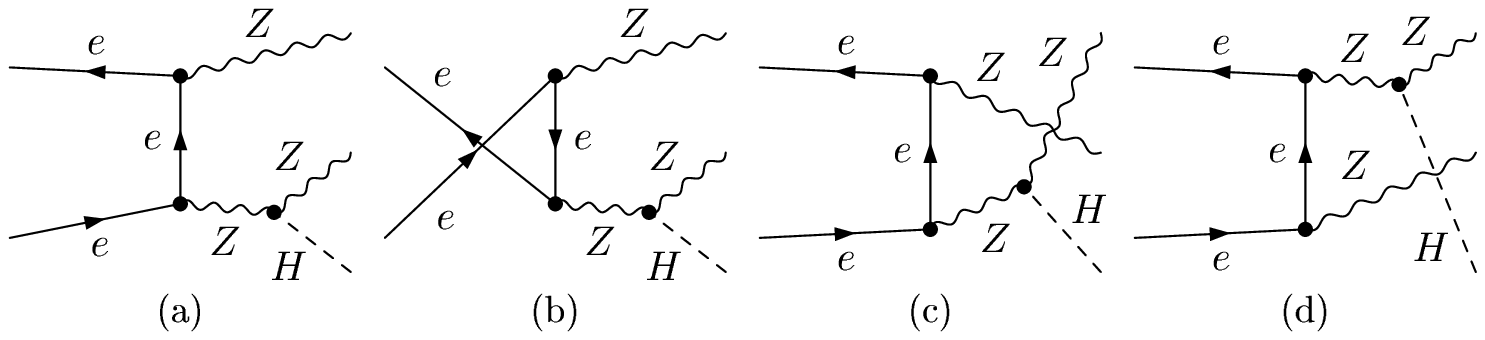}
\vspace*{-0.3cm} \centering \caption{\label{tree} The tree-level
Feynman diagrams for the process \eezzh.}
\end{figure}

The tree-level Feynman diagrams contributing to the process \eezzh
in the frame of the SM are depicted in Fig.\ref{tree}. Due to the
fact that the Yukawa coupling strength between Higgs/Goldstone and
fermions is proportional to the fermion mass, it is reasonable to
neglect the contributions of the Feynman diagrams which include
$H^0-e^+-e^-$ or $G^0-e^+-e^-$ coupling.

\par
We calculated the Born cross section of the process \eezzh by
using 't Hooft-Feynman gauge and unitary gauge to check the gauge
invariance by adopting $FeynArts~3.2$ package\cite{Feynarts}, and
got the coincident numerical results. The electroweak one-loop
Feynman diagrams can be classified into self-energy, triangle, box
and pentagon diagrams. As a representative selection, the pentagon
diagrams are depicted in Fig.2. Their corresponding amplitudes may
involve five point tensor integrals up to rank 4. In the amplitude
calculation of the process \eezzh involving one-loop
contributions, we create all the tree-level, one-loop Feynman
diagrams and their relevant amplitudes in the 't Hooft-Feynman
gauge by using $FeynArts~3.2$, and the Feynman amplitudes are
subsequently reduced by $FormCalc~4$ \cite{Form}. Our
renormalization procedure is implemented in these packages. The
numerical calculation of the two-, three- and four-point integral
functions are done by using FF package\cite{van}. The
implementations of the scalar and the tensor five-point integrals
are done exactly by using the Fortran programs as used in our
previous works on $e^+e^- \to t \bar t H^0$ and $e^+e^- \to
Z^0H^0H^0$ processes\cite{youyu,ZhangRY} with the approach
presented in Ref.\cite{pentagon}.
\begin{figure}[htbp]
\vspace*{-0.3cm} \centering
\includegraphics*[130pt,282pt][560pt,618pt]{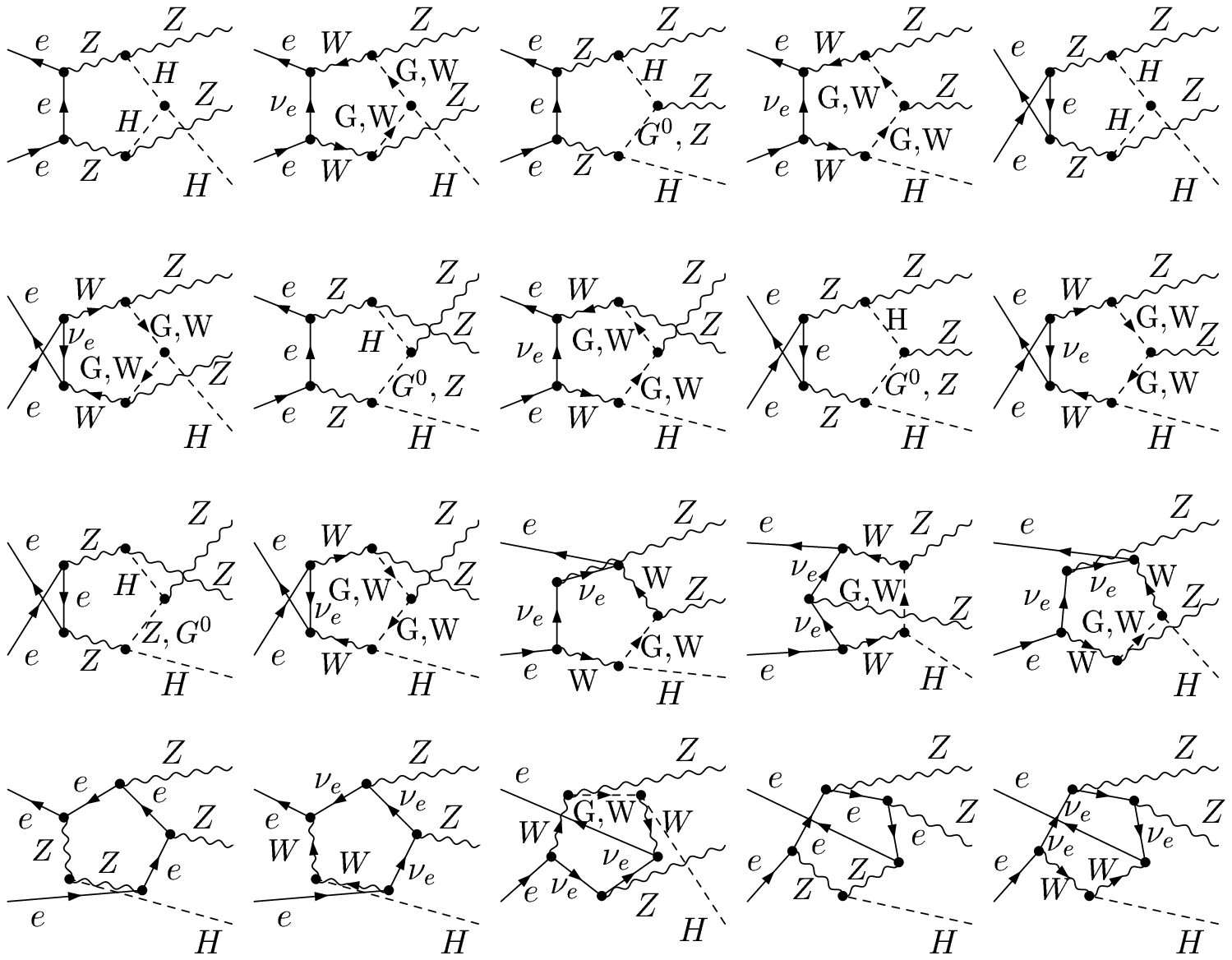}
\vspace*{-0.3cm} \centering \caption{\label{pentagon} The pentagon
Feynman diagrams for the process \eezzh.}
\end{figure}

\par
The ${\cal O}(\alpha_{ew})$ virtual electroweak correction to the
\eezzh process can be expressed as:
\begin{eqnarray}
\sigma_{\rm virtual}=\sigma_{\rm tree}\delta_{\rm virtual}=\frac{1}{2}
\frac{(2 \pi )^4}{2|\vec{p}_1|\sqrt{s}}\int d\Phi_3\overline{\sum_{spin}}
Re({\cal M}_{\rm tree}{\cal M}^{\dag}_{\rm virtual})
\end{eqnarray}
where the first factor $\frac{1}{2}$ comes from the two identical
$Z^0$-bosons in the final state, $\vec{p}_1$ is the momentum of
the incoming positron in the center of mass system(c.m.s.), ${\rm
d} \Phi_3$ is the three-body phase space element, and the bar over
summation recalls averaging over initial spins \cite{RevD}.
$\sigma_{\rm tree}$ and ${\cal M}_{\rm tree}$ are the cross
section and amplitude at the tree-level for process \eezzh,
respectively. ${\cal M}_{\rm virtual}$ is the renormalized
amplitude from all the electroweak one-loop Feynman diagrams and
the corresponding counterterms. The related renormalized
quantities and renormalization constants are defined as in
Ref.\cite{Denner}, and can be evaluated by using the corresponding
equations shown in this reference.

\par
The total unrenormalized amplitude corresponding to all the
one-loop Feynman diagrams contains both ultraviolet (UV) and
infrared (IR) divergences. To regularize the UV divergences in
loop integrals, we adopt the dimensional regularization scheme
\cite{DR} in which the dimensions of spinor and space-time
manifolds are extended to $D = 4 - 2 \epsilon$. And we adopt the
on-mass-shell (OMS) scheme \cite{COMS scheme, Denner} to
renormalize the relevant fields. All the tensor coefficients of
the one-loop integrals can be calculated by using the reduction
formulae presented in Refs.\cite{Passarino,Dittmaier}. We check
the UV finiteness of our results of the whole contributions of the
virtual one-loop diagrams and counterterms both analytically and
numerically by regularizing the IR divergence with a fictitious
photon mass. As we expect, the UV divergence contributed by
virtual one-loop diagrams can be cancelled by that contributed
from the counterterms exactly.

\par
The soft IR divergence in the process \eezzh is originated from
virtual photonic corrections, which can be exactly cancelled by
adding the real photonic bremsstrahlung corrections to this
process in the soft photon limit. In the real photon emission
process
\begin{eqnarray}
\label{real photon emission}
 e^+(p_1)+e^-(p_2) \rightarrow Z^0(k_1)+Z^0(k_2)+H^0(k_3)+\gamma(k_{\gamma}),
\end{eqnarray}
a real photon radiates from the electron/positron, and can have
either soft or collinear nature. The collinear singularity is
regularized by keeping electron(positron) mass. We use the general
phase-space-slicing (PSS) method \cite{PSS} to isolate the soft
photon emission singularity in the real photon emission process.
In the PSS method, the bremsstrahlung phase space is divided into
singular and non-singular regions, and the cross section of the
real photon emission process ($\ref{real photon emission}$) is
decomposed into soft and hard terms
\begin{equation}
\sigma_{{\rm real}}=\sigma_{{\rm soft}}+\sigma_{{\rm hard}}=
\sigma_{\rm tree}(\delta_{{\rm soft}}+\delta_{{\rm hard}}).
\end{equation}
where the 'soft' and 'hard' describe the energy of the radiated
photon. The energy $E_{\gamma}$ of the radiated photon in the
center of mass system(c.m.s.) frame is considered soft and hard if
$E_{\gamma} \leq \Delta E$ and $E_{\gamma} > \Delta E$,
respectively. Both $\sigma_{{\rm soft}}$ and $\sigma_{{\rm hard}}$
depend on the arbitrary soft cutoff $\Delta E/E_{b}$, where $E_b =
\sqrt{s}/2$ is the electron beam energy in the c.m.s. frame, but
the total cross section of the real photon emission process
$\sigma_{{\rm real}}$ is cutoff independent. Since the soft cutoff
$\Delta E/E_{b}$ is taken to be a small value in our calculations,
the terms of order $\Delta E/E_{b}$ can be neglected and the soft
contribution can be evaluated by using the soft photon
approximation analytically \cite{COMS scheme,Denner, soft r
approximation}
\begin{eqnarray}
\label{soft part} {\rm d} \sigma_{{\rm soft}} = -{\rm d}
\sigma_{{\rm tree}}  \frac{\alpha_{ew}}{2 \pi^2}
\int_{|\vec{k}_{\gamma}| \leq \Delta E} \frac{{\rm d}^3 k_{\gamma}}{2
E_{\gamma}} \left( \frac{p_1}{p_1 \cdot k_{\gamma}} -
\frac{p_2}{p_2 \cdot k_{\gamma}} \right)^2.
\end{eqnarray}
As shown in Eq.(\ref{soft part}), the soft contribution has an IR
singularity at $m_{\gamma} = 0$, which can be cancelled exactly
with that from the virtual photonic corrections. Therefore,
$\sigma_{{\rm virtual} + {\rm soft}}$, the sum of the ${\cal
O}(\alpha_{ew}^4)$ virtual and soft photon emission cross section
corrections, is independent of the fictitious small photon mass
$m_{\gamma}$. The hard contribution, which is UV and IR finite, is
computed by using the Monte Carlo technique. Finally, the
corrected cross section for the \eezzh process up to the order of
${\cal O}(\alpha^4_{ew})$ is obtained by summing the ${\cal
O}(\alpha^3_{ew})$ Born cross section $\sigma_{\rm tree}$, the
${\cal O}(\alpha^4_{ew})$ virtual cross section $\sigma_{\rm
virtual}$, and the ${\cal O}(\alpha^4_{ew})$ cross section of the
real photon emission process (\ref{real photon emission}), i.e.,
\begin{eqnarray}
\sigma_{{\rm total}}= \sigma_{{\rm tree}} + \sigma_{{\rm virtual}}
+ \sigma_{{\rm real}} = \sigma_{{\rm tree}} \left( 1 +
\delta_{{\rm total}} \right),
\end{eqnarray}

\par
In order to analyze the origins of the one-loop electroweak
corrections clearly, we organize the full one-loop electroweak
corrections to the process \eezzh in two parts, the QED correction
part and the weak correction part. The QED part comes from the QED
virtual correction and the real photon emission correction. The
QED virtual correction involves the contributions from the
one-loop diagrams with virtual photon exchange in the loop. For
some counterterms involved in the QED contribution, we only have
to take into account purely photonic contribution to the wave
function renormalization constants of the electron/positron. The
rest of the total virtual electroweak corrections is called the
weak correction part. With such definitions of the origins of the
radiative correction we can divide the full one-loop electroweak
corrected total cross section in following form:
\begin{eqnarray}
\sigma_{{\rm total}}&=& \sigma_{{\rm tree}} + \sigma_{{\rm
virtual+soft}} + \sigma_{{\rm hard}}= \sigma_{{\rm tree}} +
\sigma^{QED}_{{\rm virtual+soft}} +
\sigma^{QED}_{{\rm hard}} + \sigma^{W} \nb\\
&=& \sigma_{{\rm tree}} \left( 1 + \delta^{QED}+\delta^{W}
\right)= \sigma_{{\rm tree}} \left( 1 + \delta_{{\rm total}}
\right),
\end{eqnarray}
where $\sigma_{{\rm virtual+soft}}$ is the cross section
correction contributed by the virtual electroweak one-loop
diagrams and the soft photon emission process, $\sigma^{QED}_{{\rm
virtual+soft}}$, $\sigma^{QED}_{{\rm hard}}$ and $\sigma^{W}$ are
the corrections from the QED contributions including only the
photonic one-loop diagrams and the soft photon emission process,
the hard photon emission process and the weak virtual
contribution, separately. $\delta^{QED}_{{\rm virtual+soft}}$ is
the relative correction including the contributions of QED
one-loop diagrams and the soft photon emission process.
$\delta^{QED}$, $\delta^{W}$ and $\delta_{{\rm total}}$ are the
relative corrections contributed by the QED correction part, the
weak correction part and the total electroweak correction,
respectively.

\par
\section{Numerical Results and Discussions}

\par
In our numerical calculation, we adopt the $\alpha_{ew}$-scheme,
and the input parameters are taken as follows\cite{RevD}:
\begin{eqnarray}
m_e&=&0.510998902~{\rm MeV},~m_\mu~=~105.658369~{\rm MeV},~m_\tau~=~1776.99~{\rm MeV},\nb\\
m_u&=&66~{\rm MeV},~~~~~~~~~~~~~~m_c~=~1.2~{\rm GeV},~~~~~~~~~~~~~m_t~=~178.1~{\rm GeV},\nb\\
m_d&=&66~{\rm MeV},~~~~~~~~~~~~~~m_s~=~150~{\rm MeV},~~~~~~~~~~~~m_b~=~4.3~{\rm GeV} ,\nb\\
m_W&=&80.425~{\rm GeV},~~~~~~~~~m_Z~=~91.1876~{\rm
GeV}.~~~~~~~\alpha_{ew}(0)=1/137.036
\end{eqnarray}
There we take the electric charge defined in the Thomson limit
$\alpha_{ew}(0) = 1/137.036$ and the effective values of the light
quark masses ($m_u$ and $m_d$) which can reproduce the hadronic
contribution to the shift in the fine structure constant
$\alpha_{ew}(m_Z^2)$ \cite{DESY}.

\par
Besides the parameters mentioned above, we must provide the values
of the regulator $m_{\gamma}$ and soft cutoff $\Delta E/E_b$. As
we know, the total cross section should have no relation with
these two parameters. We checked the photon mass independence of
the total cross section, and found that in the cases of
$m_{\gamma}=10^{-20}~{\rm GeV}$ and $m_{\gamma}=1~{\rm GeV}$, the
${\cal O} (\alpha^4_{ew})$ cross sections $\sigma_{\rm
virtual+soft}$ are invariable within the statistical error
$1.0\times 10^{-4}$. In the numerical calculation it shows that if
we take a very small value for the regulator $m_{\gamma}$, the
Monte Carlo sampling of the virtual correction part is very slow
and it will take a long time to get requested accuracy. With this
consideration we take $m_{\gamma}=1~{\rm GeV}$ and $\Delta
E/E_b=10^{-2}$, if there is no other statement.

\par
\begin{figure}[htp]
\vspace*{-0.3cm} \centering
\includegraphics[scale=0.7,bb=32 32 520 405]{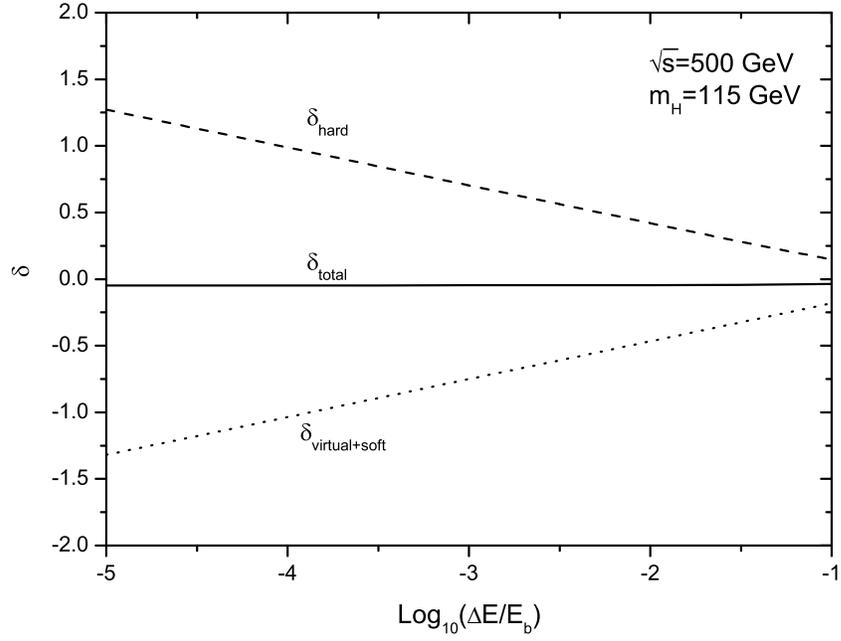}
\vspace*{-0.3cm} \centering \caption{\label{fig3} The ${\cal
O}(\alpha_{ew})$ relative correction to the \eezzh process as a
function of the soft cutoff $\Delta E/E_b$}
\end{figure}

To show the independence of the total correction on the soft
cutoff $\Delta E/E_b$, we present the ${\cal O}(\alpha_{ew})$
relative correction to the \eezzh process as a function of $\Delta
E/E_b$ in Fig.\ref{fig3}, with $m_H=115~{\rm GeV}$ and
$\sqrt{s}=500~{\rm GeV}$. As shown in this figure, both
$\delta_{\rm soft+virtual}$ and $\delta_{\rm hard}$ obviously
depend on the soft cutoff $\Delta E/E_b$, but the full ${\cal
O}(\alpha_{ew})$ electroweak relative correction $\delta_{\rm
total}$ is independent of the soft cutoff value.
\vskip 10mm

\par
\begin{figure}[hbp]
\vspace*{-0.3cm} \centering
\includegraphics[scale=0.7,bb=32 32 535 392]{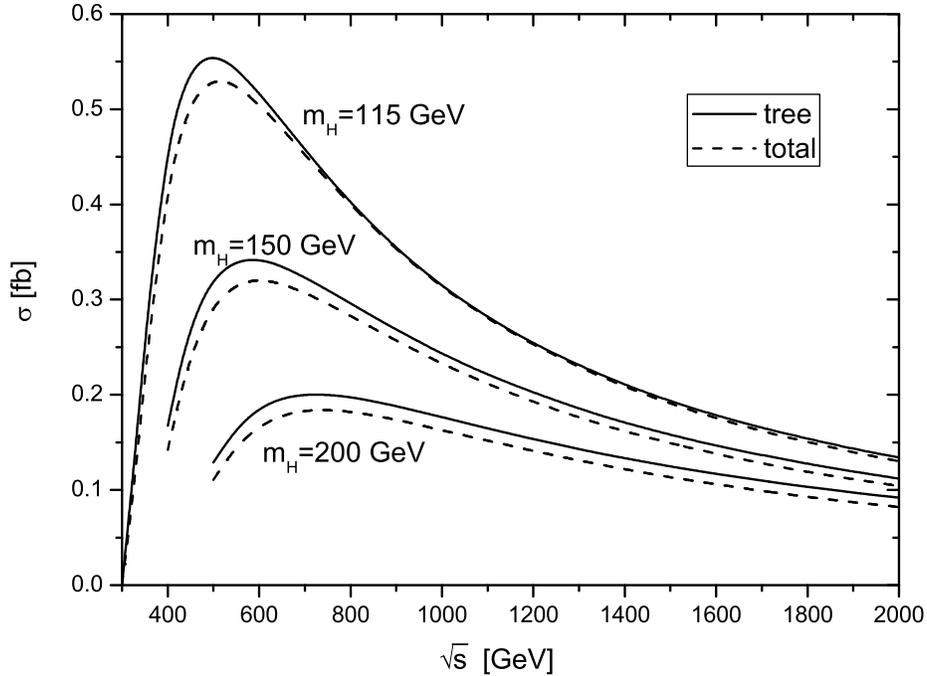}
\vspace*{-0.3cm} \centering \caption{\label{fig4} The Born and
one-loop level corrected cross sections for the \eezzh process as
the functions of the $e^+e^-$ colliding energy $\sqrt{s}$}
\end{figure}

In Fig.\ref{fig4} we present the Born cross section $\sigma_{\rm
tree}$ and the full one-loop electroweak corrected cross section
$\sigma_{\rm total}$ as the functions of the c.m.s. energy
$\sqrt{s}$ with $m_H=115~{\rm GeV},~150~{\rm GeV}$ and $200~{\rm
GeV}$, respectively. We can see that the corrected cross sections
are always less than the corresponding tree-level cross sections
clearly, i.e., the radiative corrections are always negative in
our chosen parameter spaces. We also find that the curves for both
$\sigma_{\rm tree}$ and $\sigma_{\rm total}$ go up rapidly to
reach their maximal values with the increment of $\sqrt{s}$ in the
region near the threshold, and then go down to approach small
values. We can read out from the figure that the cross sections
$\sigma_{\rm tree}$ and $\sigma_{\rm total}$ reach their maximal
values of about $0.555~fb$ and $0.529~fb$ respectively, in the
vicinity of $\sqrt{s}\sim 500~{\rm GeV}$ when $m_H=115~{\rm GeV}$.
For $m_H=150~{\rm GeV}$, the maximal values of $\sigma_{\rm tree}$
and $\sigma_{\rm total}$ are at the position around $\sqrt{s}\sim
600~{\rm GeV}$ and have the values about $0.342~fb$ and $0.320~fb$
separately. When $m_H=200~{\rm GeV}$, the $\sigma_{\rm tree}$ and
$\sigma_{\rm total}$ can reach about $0.200~fb$ and $0.184~fb$ at
$\sqrt{s}\sim 700~{\rm GeV}$, respectively.

\par
\begin{figure}[htbp]
\vspace*{-0.3cm} \centering
\includegraphics[scale=0.7,bb=32 32 527 405]{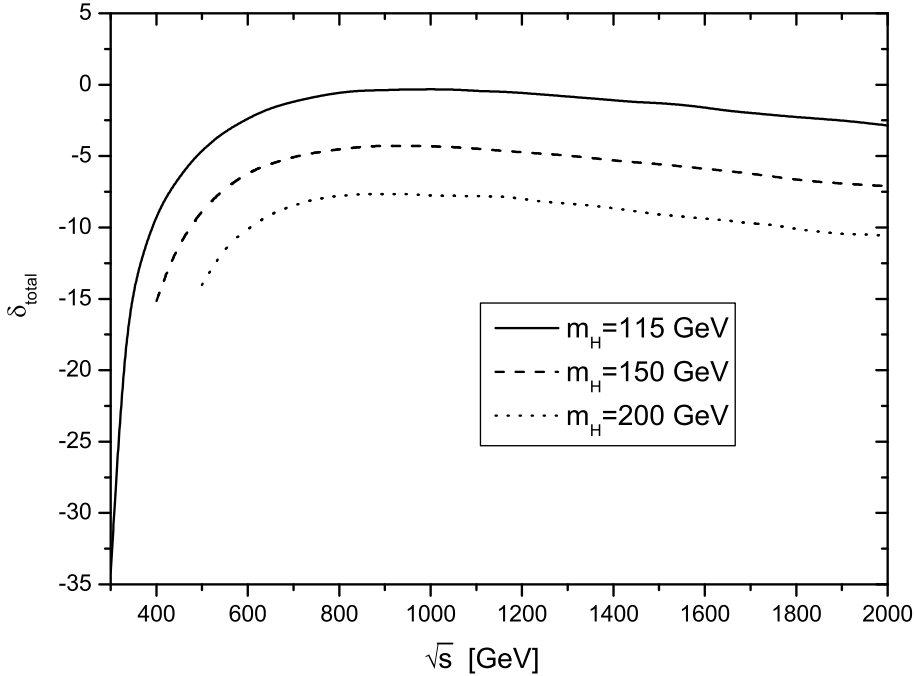}
\vspace*{-0.3cm} \centering \caption{\label{fig5} The ${\cal
O}(\alpha_{ew})$ relative corrections to the \eezzh process as the
functions of $\sqrt{s}$}
\end{figure}

Fig.\ref{fig5} displays the full ${\cal O}(\alpha_{ew})$
electroweak relative correction for the \eezzh process versus the
c.m.s. energy $\sqrt{s}$, with $m_H=115~{\rm GeV},~150~{\rm GeV}$
and $200~{\rm GeV}$, respectively. As shown in the figure, the
full ${\cal O}(\alpha_{ew})$ electroweak corrections suppress the
Born cross sections in the range of $300~{\rm GeV}\le\sqrt{s}\le
2000~{\rm GeV}$. The relative correction can be beyond $-30\%$
near $\sqrt{s}=300~{\rm GeV}$ with $m_H=115~{\rm GeV}$, but that
value near the threshold is phenomenologically insignificant. When
the c.m.s. energy is far beyond the threshold, the relative
correction becomes insensitive to $\sqrt{s}$. The values of the
relative corrections vary in the ranges from $-0.32\%$ to
$-2.88\%$, from $-4.28\%$ to $-7.14\%$, and from $-7.64\%$ to
$-10.6\%$ when $\sqrt{s}$ goes up from $800~{\rm GeV}$ to
$2000~{\rm GeV}$ for $m_H=115{\rm GeV},150~{\rm GeV}$ and
$200~{\rm GeV}$, respectively.

\par
\begin{figure}[htbp]
\vspace*{-0.3cm} \centering
\includegraphics[scale=0.7,bb=32 32 531 400]{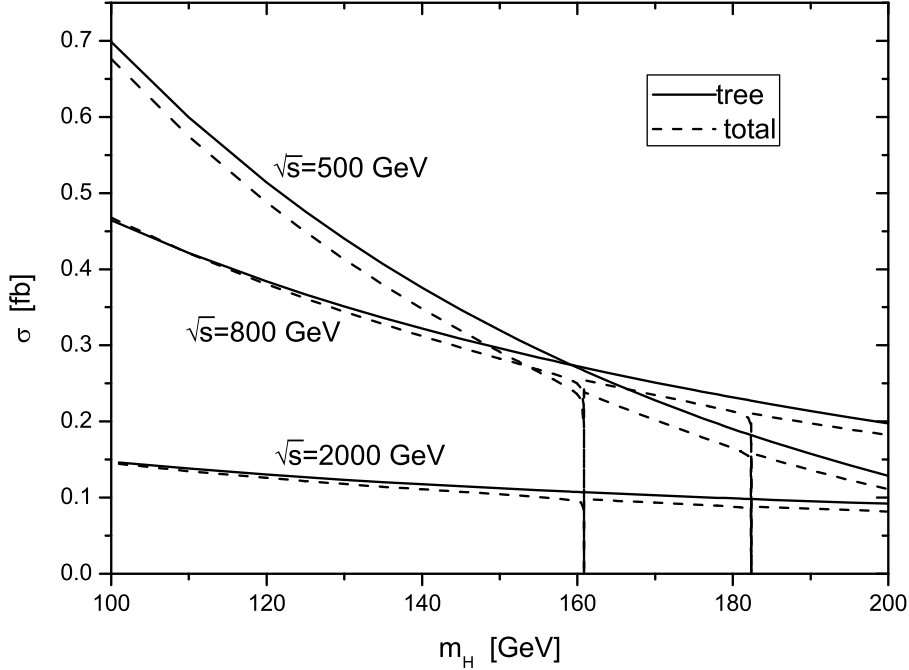}
\vspace*{-0.3cm} \centering \caption{\label{fig6} The Born cross
section and the one-loop level corrected cross section as the
functions of the Higgs-boson mass $m_H$}
\end{figure}

We plot the Born cross section $\sigma_{\rm tree}$ and the
electroweak corrected cross section $\sigma_{\rm total}$ as the
functions of the Higgs-boson mass $m_H$ in Fig.\ref{fig6} with
different values of $\sqrt{s}$. It shows that both $\sigma_{\rm
tree}$ and $\sigma_{\rm total}$ decrease with the increment of the
Higgs-boson mass, and the less the value of $\sqrt{s}$, the more
rapidly they drop. We can see in this figure that the cross
sections of $\sigma_{\rm tree}$ and $\sigma_{\rm total}$ with
$\sqrt{s}=2000~{\rm GeV}$ are insensitive to $m_H$, and the
corresponding curves are stable as the increment of $m_H$ from
$100~GeV$ to $200~GeV$. The figure shows also that each curve for
the one-loop corrected cross section has two spikes, which just
reflect the resonance effects of the virtual corrections at the
positions of $m_H=2m_W$ and $m_H=2m_Z$, respectively. Since we do
not take the complex masses of the $W^{\pm}$ and $Z^0$ bosons in
the calculation of one-loop integrals, the numerical results in
the vicinities of $m_H=2m_W$ and $m_H=2m_Z$ are not reliable.

\par
\begin{figure}[htbp]
\vspace*{-0.3cm} \centering
\includegraphics[scale=0.7,bb=32 32 527 395]{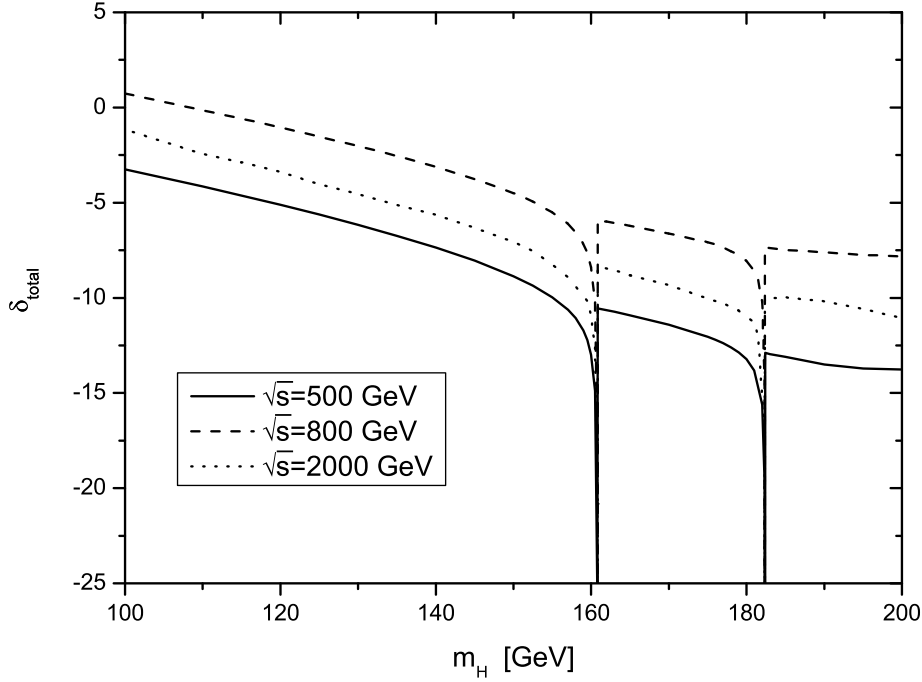}
\vspace*{-0.3cm} \centering \caption{\label{fig7} The full ${\cal
O}(\alpha_{ew})$ relative corrections to the \eezzh process as the
functions of the Higgs-boson mass $m_H$}
\end{figure}

Fig.\ref{fig7} shows the relationship between the full ${\cal
O}(\alpha_{ew})$ relative correction of the $e^+e^-\to Z^0Z^0H^0$
process and the Higgs-boson mass $m_H$. We find that the curves
for different values of colliding energy decrease with the
increment of Higgs-boson mass $m_H$. The relative corrections for
$\sqrt{s}=500~{\rm GeV}$ have negative values, and are generally
smaller than the corresponding ones for $\sqrt{s}=800~{\rm GeV}$
and $\sqrt{s}=2000~{\rm GeV}$. We can see from the figure that on
each curve there are also two spikes at the positions of $m_H=2
m_W$ and $m_H=2 m_Z$ due to the resonance effects. Again the
numerical relative corrections in those positions are unreliable
for the same reason as declared for Fig.\ref{fig6}. For
$\sqrt{s}=500~{\rm GeV}$, the most favorable colliding energy for
\eezzh process with intermediate Higgs-boson mass, the relative
correction decreases from $-3.24\%$ to $-13.8\%$ as $m_H$
increases from $100~GeV$ to $200~GeV$. We can also read out from
the figure that when Higgs-boson mass $m_H$ increases from
$100~GeV$ to $200~GeV$, the relative correction for
$\sqrt{s}=800~{\rm GeV}$ goes down from $0.728\%$ to $-7.82\%$,
and from $-1.29\%$ to $-11.1\%$ for $\sqrt{s}=2000~{\rm GeV}$.


\par
In Table 1 we list and compare some numerical results of the
radiative corrections to \eezzh process contributed by the QED
correction part, the weak correction part and the total
electroweak one-loop correction. In our calculation we set the
soft cutoff $\Delta E/E_b$ being $10^{-2}$. From the table we can
see that when $\sqrt{s}=500~GeV$ and $m_H=115,~150~GeV$, the full
QED one-loop corrections have negative signs and are much larger
than the weak corrections, while when $\sqrt{s}=1000~GeV$ and
$m_H=115,~150~GeV$, the weak corrections become important and
their absolute values of the weak correction part are comparable
with those of the QED correction part. For the case with
$\sqrt{s}=1000~GeV$ and $m_H=115~GeV$, the large cancellation
between the corrections from the QED and weak parts makes a
relative small total one-loop electroweak correction.

\begin{table}[htbp]
\begin{center}
\begin{tabular}{c|c|c|c|c|c|c|c}
\hline $\sqrt{s}$(GeV)& $m_H$(GeV) & $ \sigma^{QED}_{V+S}(fb)$ & $
\sigma^{QED}_{hard} (fb)$ & $ \sigma^W(fb)$ &
$\delta_{total}$(\%)& $\delta^{QED}$(\%) &
$\delta^W$(\%) \\
\hline
500 & 115 & -0.2587 & 0.2333 & -0.0002 & -4.621 & -4.576 & -0.045 \\[0mm]
    & 150 & -0.1491 & 0.1223 & -0.0014 & -8.857 & -8.412 & -0.445 \\[0mm]
\hline
1000& 115 & -0.1538 & 0.1738 & -0.0210 & -0.320 &  6.360 & -6.680  \\[0mm]
    & 150 & -0.1191 & 0.1293 & -0.0207 & -4.317 &  4.204 & -8.521  \\[0mm]
\hline
\end{tabular}
\caption{The comparison between the numerical results contributed
by the QED correction part, the weak correction part and the total
electroweak ${\cal O}(\alpha_{ew})$ correction to \eezzh process}
\end{center}
\end{table}

\section{Summary}
In this paper we calculate the full ${\cal O}(\alpha_{ew})$
electroweak radiative correction to the $e^+e^-\to Z^0Z^0H^0$
process within the framework of the SM at linear colliders. We
analyze the dependence of the Born cross section, the full ${\cal
O}(\alpha_{ew})$ electroweak corrected cross section and the
relative correction on colliding energy $\sqrt{s}$ and Higgs-boson
mass $m_H$. From the numerical results we find that the full
${\cal O}(\alpha_{ew})$ electroweak correction significantly
suppresses the Born cross section. When $\sqrt{s}=2000~{\rm GeV}$,
both the Born and corrected cross sections are insensitive to
Higgs-boson mass $m_H$. But when $\sqrt{s}$ is relatively smaller,
e.g., $\sqrt{s}=500~{\rm GeV}$ or $800~{\rm GeV}$, both the Born
and corrected cross sections decrease sharply with the increment
of Higgs-boson mass $m_H$. In our chosen parameter space, the
relative corrections are in the value range between $1.0\%$ and
$-15\%$. These corrections are so remarkable that we must consider
them in the precise experimental analysis.

\vskip 10mm
\noindent{\large\bf Acknowledgments:}
\par
This work was supported in part by the National Natural Science
Foundation of China and a special fund sponsored by Chinese
Academy of Sciences.

\newpage
\vskip 10mm
\begin{flushleft} {\bf Figure Captions} \end{flushleft}
\par
{\bf Figure 1} The tree-level Feynman diagrams for the process \eezzh.

\par
{\bf Figure 2} The pentagon Feynman diagrams for the process \eezzh.

\par
{\bf Figure 3} The ${\cal O}(\alpha_{ew})$ relative correction to
the \eezzh process as a function of the soft cutoff $\Delta E/E_b$.

\par
{\bf Figure 4} The Born and one-loop level corrected cross sections
for the \eezzh process as the functions of the $e^+e^-$ colliding
energy $\sqrt{s}$.

\par
{\bf Figure 5} The ${\cal O}(\alpha_{ew})$ relative corrections to
the \eezzh process as the functions of $\sqrt{s}$.

\par
{\bf Figure 6} The Born cross section and the one-loop level
corrected cross section as the functions of the Higgs-boson mass
$m_H$.

\par
{\bf Figure 7} The full ${\cal O}(\alpha_{ew})$ relative
corrections to the \eezzh process as the functions of the
Higgs-boson mass $m_H$.

\end{document}